\documentclass[11pt, aip, prl, reprint, superscriptaddress]{revtex4-1}

\usepackage{graphicx}
\usepackage{amsmath,amssymb}
\usepackage{ulem}
\usepackage{color}
\usepackage{gensymb}
\usepackage{appendix}

\begin{document}

\title{Subwavelength pulse focusing and perfect absorption in the Maxwell fisheye}

\author{Gautier Lefebvre}
\affiliation{Universit\'e de technologie de Compi\`egne, Roberval (Mechanics, energy and electricity), Centre de recherche Royallieu - CS 60319 - 60203 Compi\`egne Cedex - France}	

\author{Marc Dubois}
\affiliation{
Aix-Marseille Univ, CNRS, Centrale Marseille, Institut Fresnel, 
13013 Marseille, France
\textcolor{black}{\& Multiwave Imaging SAS, Marseille 13013, France}
}
\author{Younes Achaoui}
\affiliation{
Laboratory for the Study of Advanced Materials and Applications, Department of Physics, Moulay Ismail University, B.P. 11201, Zitoune, Meknes, Morocco
\\
}
\author{Ros Kiri Ing}
\affiliation{
Institut Langevin, ESPCI ParisTech CNRS UMR7587, 1 rue Jussieu, 75238 Paris cedex 05, France\\
}
\author{Mathias Fink}
\affiliation{
Institut Langevin, ESPCI ParisTech CNRS UMR7587, 1 rue Jussieu, 75238 Paris cedex 05, France\\
}

\author{S\'ebastien Guenneau}
\affiliation{
UMI 2004 Abraham de Moivre-CNRS, Imperial College London, London SW7 2AZ, United Kingdom\\
}
\affiliation{
The Blackett Laboratory, Department of Physics, Imperial College London, London SW7 2AZ, United Kingdom\\
}
\author{Patrick Sebbah}
\email[Contact: ]{patrick.sebbah@biu.ac.il}
\affiliation{
Institut Langevin, ESPCI ParisTech CNRS UMR7587, 1 rue Jussieu, 75238 Paris cedex 05, France\\
}
\affiliation{Department of Physics, The Jack and Pearl Resnick Institute for Advanced Technology, Bar-Ilan University, Ramat-Gan, 5290002 Israel}

\date{\today}

\begin{abstract}
Maxwell's fisheye is a paradigm for an absolute optical instrument with a refractive index deduced from the stereographic projection of a sphere on a plane. We investigate experimentally the dynamics of flexural waves in a thin plate with a thickness varying according to the Maxwell fisheye index profile and a clamped boundary. We demonstrate subwavelength focusing and temporal pulse compression at the image point. This is achieved by introducing a sink emitting a cancelling signal optimally shaped using a time-reversal procedure. Perfect absorption and outward going wave cancellation at the focus point are demonstrated. The time evolution of the kinetic energy stored inside the cavity reveals that the sink absorbs energy out of the plate ten times faster than the natural decay rate. 

\end{abstract}
\maketitle

All rays originating from any object point meet again at a single image point. This definition of an absolute optical instrument has been first embodied by the Maxwell fisheye \cite{maxwell1854}. By means of a proper spatial variation of the refractive index, all rays emanating from a source at any position will refocus at another image point. This special class of instruments has attracted strong interest for the purpose of perfect imaging. Since the discovery of Maxwell, the family of absolute instruments has vastly expanded. The development of transformation optics, based on geometric transformation of space, has provided a new way to design such devices \cite{Tyc2011}. Among them are Luneburg or Eaton lenses, which have been revisited by this strategy. Beyond the paradigm of absolute instruments, transformation optics has proven to be a very efficient way of manipulating light for a variety of purposes, including invisibility cloaks \cite{Pendry2006, Leonhardt2006,leonhardt2009broadband}. Recently, a different and more general approach on absolute instruments has been proposed, based on the Hamilton-Jacobi equation\cite{Tyc2017}, giving access to a broader variety of absolute instruments. 

The transformation strategy implies manufacturing the distribution of physical properties of the propagating medium itself, e.g. permittivity and permeability (hence refractive index) in the case of optics. Recently, a group managed to fabricate a fisheye lens integrated on a silicon-on-insulator substrate, using electron-beam lithography \cite{Bitton2018}. Still, such realizations are not easy to achieve in practice. In the field of elastic waves where the principles of transformation optics can also be applied, the propagation relies on mechanical properties of the medium, which can be designed appropriately more easily than in optics \cite{climente2014gradient, Lefebvre2015, Tang2021}. The focusing properties of the Maxwell's fisheye have been demonstrated experimentally for flexural waves propagating in a thin plate, where spatially-dependent phase velocity is obtained by varying the plate thickness \cite{Lefebvre2015}. However, as the fisheye cannot extend to infinity in practice, perfect imaging could not be achieved: part of the wavefield is lost and cannot converge to the image point. This limitation can be circumvented by closing the fisheye with mirror boundaries as proposed in \cite{Leonhardt2009}, and a real absolute instrument can be achieved in the cavity. 

In the closed Maxwell fisheye, perfect imaging is guaranteed from the viewpoint of geometrical optics. However, when considering a wavefield, the actual resolution is limited by diffraction. Unless an absorbing mechanism is introduced, resolution is limited to half a wavelength due to the interference between the converging and diverging waves at the focal point \cite{Cassereau1992}. This subtle point has been at the heart of a now settled controversy \cite{Leonhardt2009,Blaikie2010,Leonhardt2010,guenneau2010,sun2010can,Merlin2011,he2015can}. An active absorber has been proposed in a theoretical and numerical study providing the optimal drain signal, \cite{Tyc2014} that can achieve broadband sub-wavelength focusing at the image point of the Maxwell fisheye. Still, to the authors' knowledge, this has never been confirmed experimentally. 

\begin{figure}[]
(a)
\includegraphics[width=6.5cm]{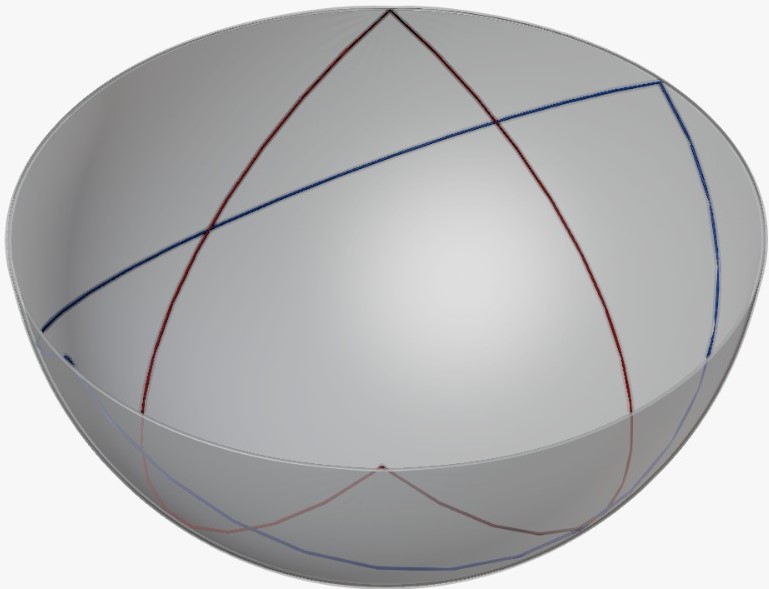}

(b)
\includegraphics[width=7.3cm]{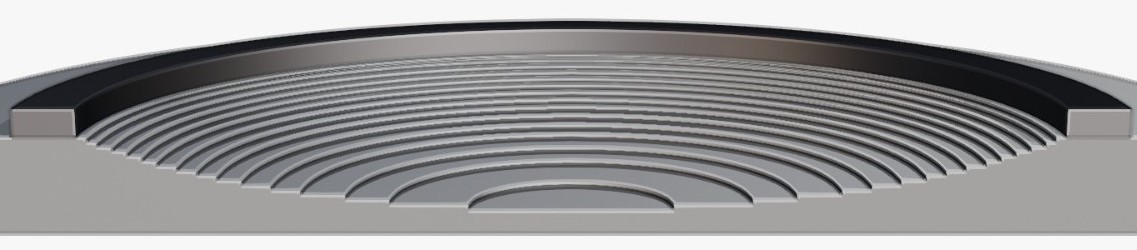}
%\mbox{}\vspace{-4cm}
%\mbox{}\hspace{-20cm}

(c)\includegraphics[width=8cm]{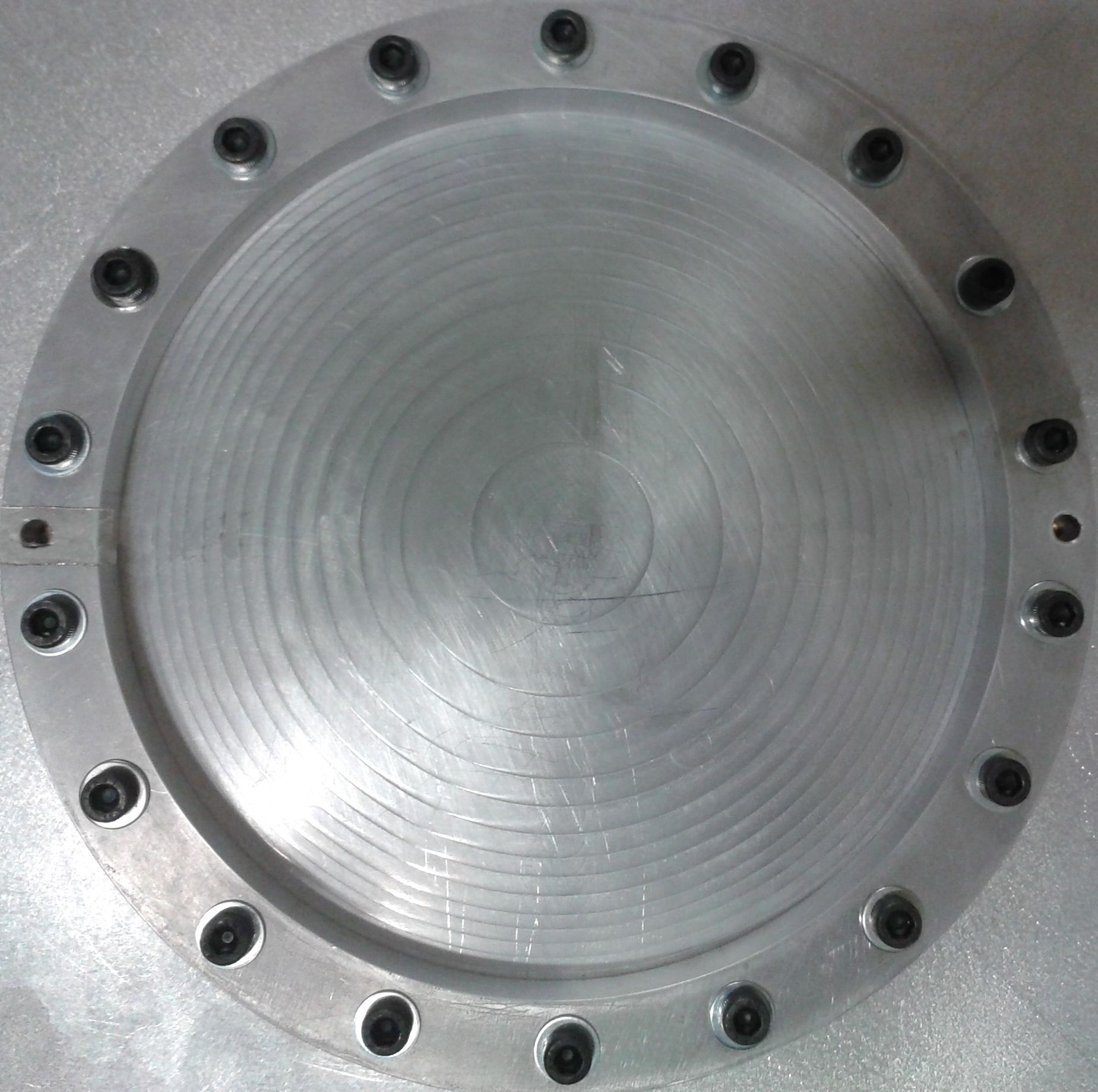}

\caption{Ray trajectories on the half-sphere, 
between two symmetrical points, considering a mirror reflection at the equator (a). 3D model of the Maxwell fisheye elastic lens, sectional view  (b); Photo of the experimental device, top view (c). The circular cavity is closed by screwed aluminium rings with inner diameter of 20$~$cm. .
}
\label{fig_intro}
\end{figure}

A very different approach to obtain point-to-point focusing is to consider a time-reversal (TR) operation in a chaotic cavity. Twenty five years ago, it was  demonstrated experimentally that ergodicity in a chaotic cavity provides refocusing of a time-reversed elastic pulse both in time and space, with a single detector in the cavity in place of a time reversal mirror on the boundary  \cite{draeger1997one,Draeger1999}. Because spatial refocusing with TR remains diffraction-limited, a time-reversed source was introduced at focus point as a sink to cancel the after-focusing outgoing wave, resulting in the sharp refocusing of the in-going wave \cite{DeRosny2002}. More recently, a passive absorber demonstrated sub-wavelength focusing with a similar time-reversal procedure in a homogeneous space \cite{ma2018}, but only with a limited bandwidth. In any case, the temporal refocusing is not perfect: residual temporal echoes persist, even for very long time-reversed records.

In this letter, we propose an absolute instrument in the sense of waves, which means not diffraction-limited. Our aim is to to shape the optimal cancelling signal that can eliminate any after-focusing echoes. We show that the sink of a TR experiment can play this role. Taking advantage of the closed fisheye cavity, we demonstrate how the combination of a short time-reversal  signal and a sink makes it possible to cancel dynamically the outward-going wave and achieve both perfect spatial and time focusing. When associated with the Maxwell fisheye, the sink appears to act as a spectacular absorbing device, eliminating very quickly most of the field within the cavity, opening a new route toward perfect absorption.

The properties of the Maxwell fisheye lens have been long known \cite{maxwell1854}. It is obtained by the stereographic projection of a sphere on a plane, which provides in turn the following refractive index distribution to ensure focusing\cite{luneburg1966mathematical}: 
\begin{equation}\label{eq_index}
	n(r)= \frac{2} {1+(r/R_{\text{max}})^2} \; ,
\end{equation}
where $R_\text{max}$ is the radius of the sphere and $r$ the Euclidean distance from its center.

To obtain the corresponding velocity profile for flexural waves, we manufacture a plate with the adequate thickness distribution, according to the following law: 

\begin{equation}\label{eq_thickness}
%	h(r)=h_0 [1+(r/R_{\text{max}})^2]^2 \; ,
h(r)={4h_0}/{n^2(r)} \; ,
\end{equation}
with $h_0$ the minimum plate thickness at the center and $n$ the index profile given by Eq. \ref{eq_index}, in agreement with the theory of transformation elasticity (see Supplementary Material for more details). The symmetry of revolution of the profile ensures that any point is the image of its symmetric with respect to the center of the lens.
However, to achieve full angular aperture and perfect imaging in a finite-size lens, a circular reflecting boundary must be added at $r=R_\text{max}$ as suggested by Leonhardt \cite{Leonhardt2009} and depicted in fig$~$\ref{fig_intro} (a). 
The elastic Maxwell fisheye is obtained by carving 15 disks of increasing depth in a duraluminium plate (see fig.$~$\ref{fig_intro} (b-c)). The resulting plate thickness ranges from 0.5$~$mm at the center to 2$~$mm at the border. A circular clamp is added to the plate around the lens. These heavy aluminium rings play the role of mirror boundaries at the edge, ensuring that all flexural waves are reflected back. 

The source is a 1 cm-diameter piezoelectric ceramic disk (PKS1-4A1 MuRata Shock Sensor) bound to the plate using Phenyl salicylate (Salol-melting point of 43.8\degree C) to ensure rigid coupling. A laser vibrometer (Polytec sensor head OFV534 with controller OFV2500) is scanned over the flat side of the plate with a 2.4$~$mm step grid to map the spatio-temporal velocity field.

\begin{figure}[]
\begin{center}
\includegraphics[width=8cm]{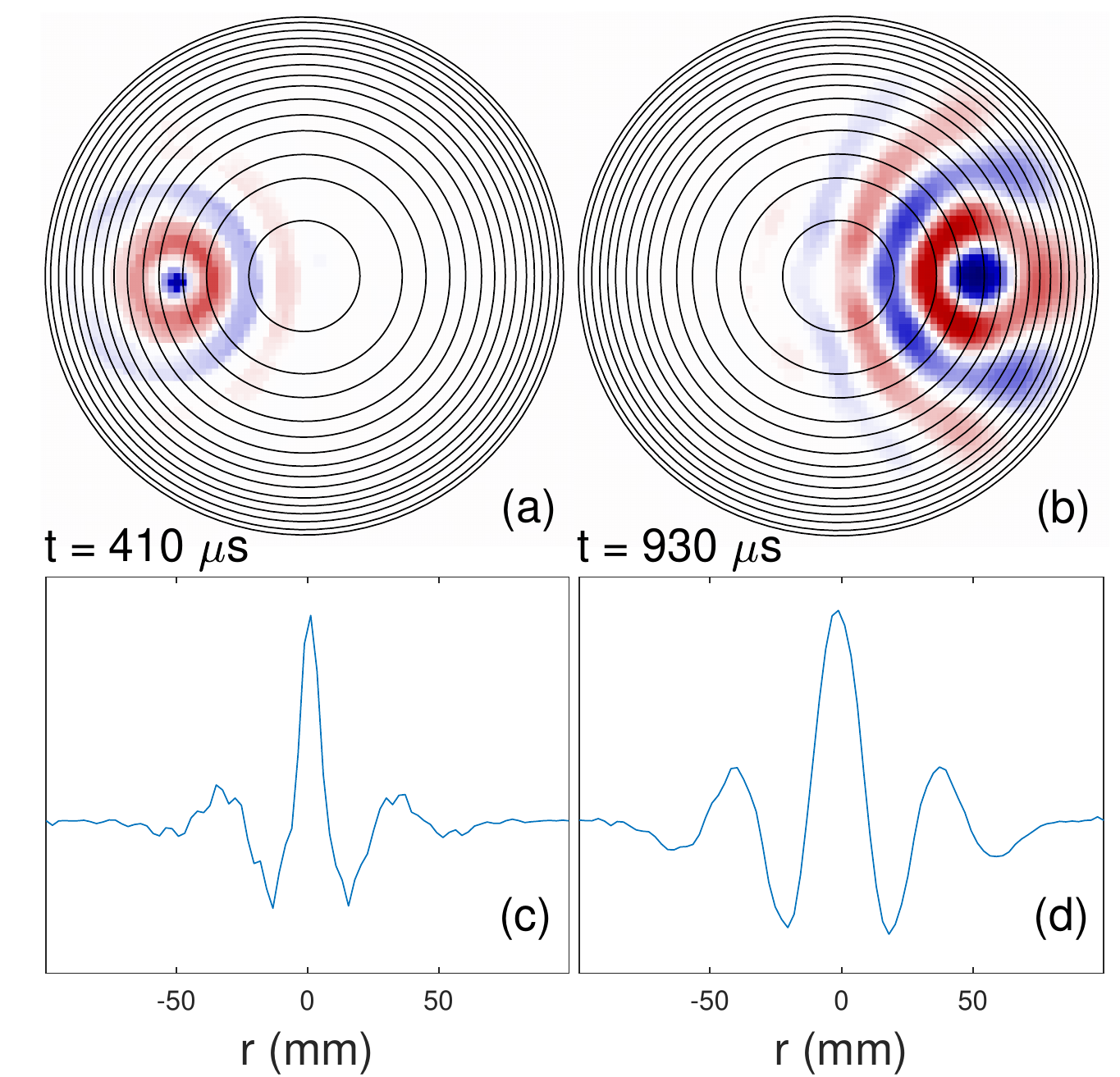}
\end{center}
%\mbox{}\vspace{-4cm}
%\mbox{}\hspace{-20cm}
\caption{Focusing of a 5kHz short pulse emitted by a single source. Upper panel displays the wavefield at emission time (a) and at focusing time (b) of the pulse. The resolution of focal spot is limited by diffraction, as visible on the lower panel showing the amplitude in arbitrary unit versus the transverse distance from the source (c) and focusing point (d). }
\label{fig_SimpleFocus}
\end{figure}

We first investigate the focusing properties of the closed Maxwell fisheye lens. A 5$~$kHz wave-packet with a 10$~$kHz bandwidth is emitted at point A, located at a radial distance of 5.3 cm  from the center of the plate. Propagation of the wavefield is monitored until focusing and beyond. The wavefield amplitude distribution measured at the maximum of the emitted pulse and of the focused pulse are shown in Fig.$~$\ref{fig_SimpleFocus}$~$a-b together with their transverse profiles taken at the source and image points Fig.$~$\ref{fig_SimpleFocus}$~$c-d. %Below, the vertical profiles at the emission and focusing point are plotted, at the same times.
The full width at half maximum (FWHM) of the emission central spot is 8.5 mm, close to the size of the piezoelectric disk. In contrast, it is 15.4 mm at focal point, almost twice the size observed at emission. At 5$~$kHz, the wavelength is 37$~$mm for the plate thickness at measurement positions. The resolution achieved is $0.42\lambda$ close to the theoretical limit of diffraction, $\lambda/2$. Although the image results from all-angle focusing, the lens is not an absolute instrument for waves and cannot image the piezoeletric disk, the size of which is smaller than the wavelength. 
This confirms that the Maxwell fisheye cannot provide sub-wavelength focusing by itself, see \cite{Leonhardt2009,Blaikie2010,Leonhardt2010,guenneau2010,Merlin2011} for a now settled controversy. Super-resolution can be achieved only with an absorbing mechanism to cancel the diverging component of the wavefield at focal point \cite{Blaikie2010, Merlin2011, Leonhardt2015}.  

\begin{figure}[]
\begin{center}
\includegraphics[width=8.5cm]{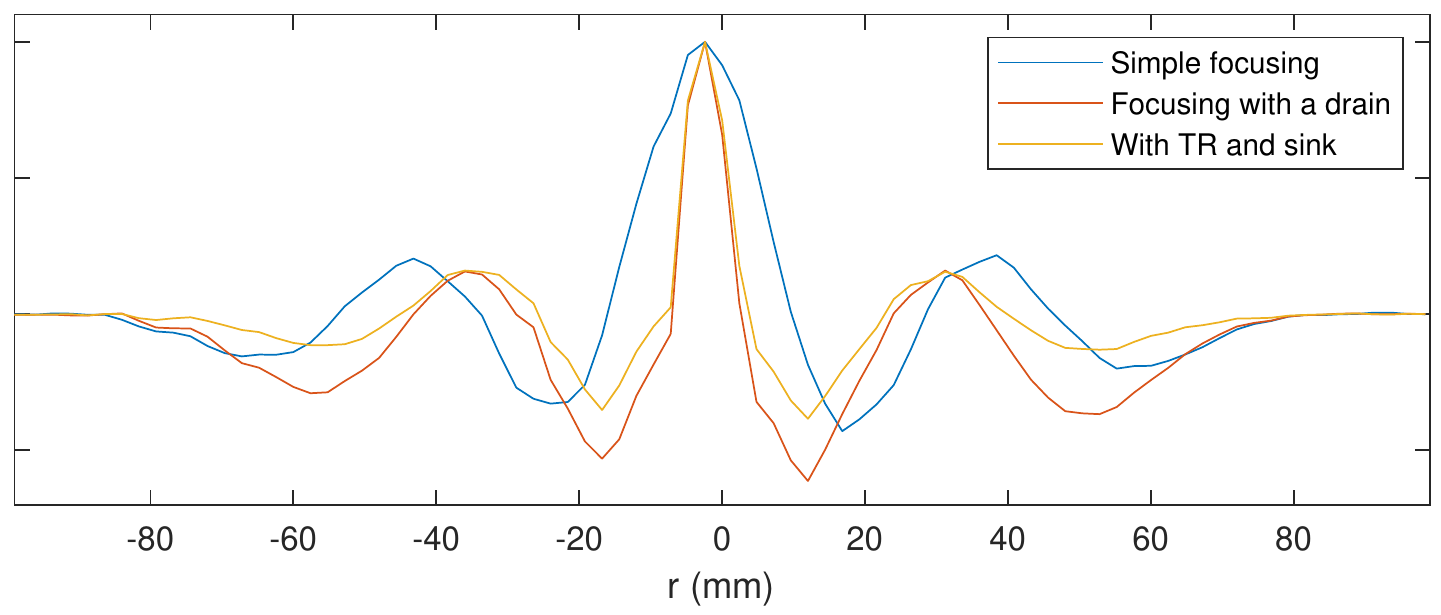}
\end{center}
\mbox{}\vspace{-1cm}
\caption{Spatial profiles of the plate vibration at focusing time, as a function of the distance from the focusing point in the vertical direction (perpendicular to the A-B axis). The amplitudes are normalized by the maximal value at $r=0$. }
\label{fig_spatialFocus}
\end{figure}

\begin{figure}[]
\begin{center}
\includegraphics[width=8.5cm]{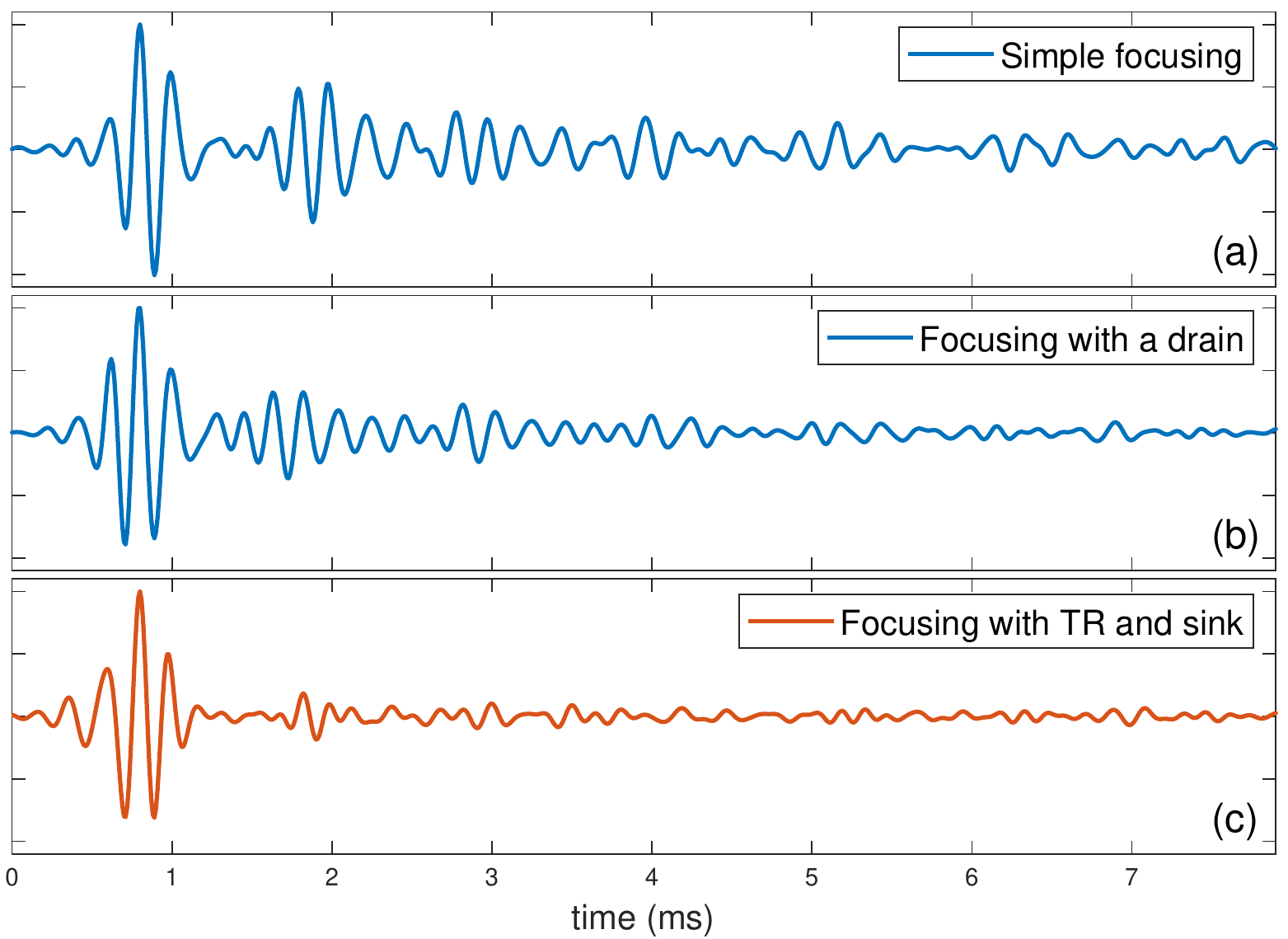}
\end{center}
\mbox{}\vspace{-1cm}
\caption{ Temporal signal in arbitrary units measured at the focusing point without use of a sink (a), in the presence of a drain (b), and using time-reversal and sink (c). }
\label{fig_timeFocus}
\end{figure}

An active drain is implemented at the image point B by introducing a source, $S_2$, which emits the opposite of the initial pulse at refocusing time. The results show that this is actually enough to obtain sub-wavelength refocusing (Fig.$~$\ref{fig_spatialFocus}). The spatial width (FWHM) of the refocused pulse spatial profile is 9$~$mm, close to the dimension of the initial source, $S_1$ (Fig. \ref{fig_SimpleFocus}c). 
However, this drain is not able to fully cancel the outgoing wavefield at focusing point. The velocity field at this position exhibits a long tail associated with successive echoes (as shown in Fig.$~$\ref{fig_timeFocus}b,  comparable to the reverberating signal observed without drain (fig.$~$\ref{fig_timeFocus}a). 

\begin{figure}[]
\begin{center}
%\mbox{}\vspace{-4cm}
%\mbox{}\hspace{-20cm}
\includegraphics[width=9cm]{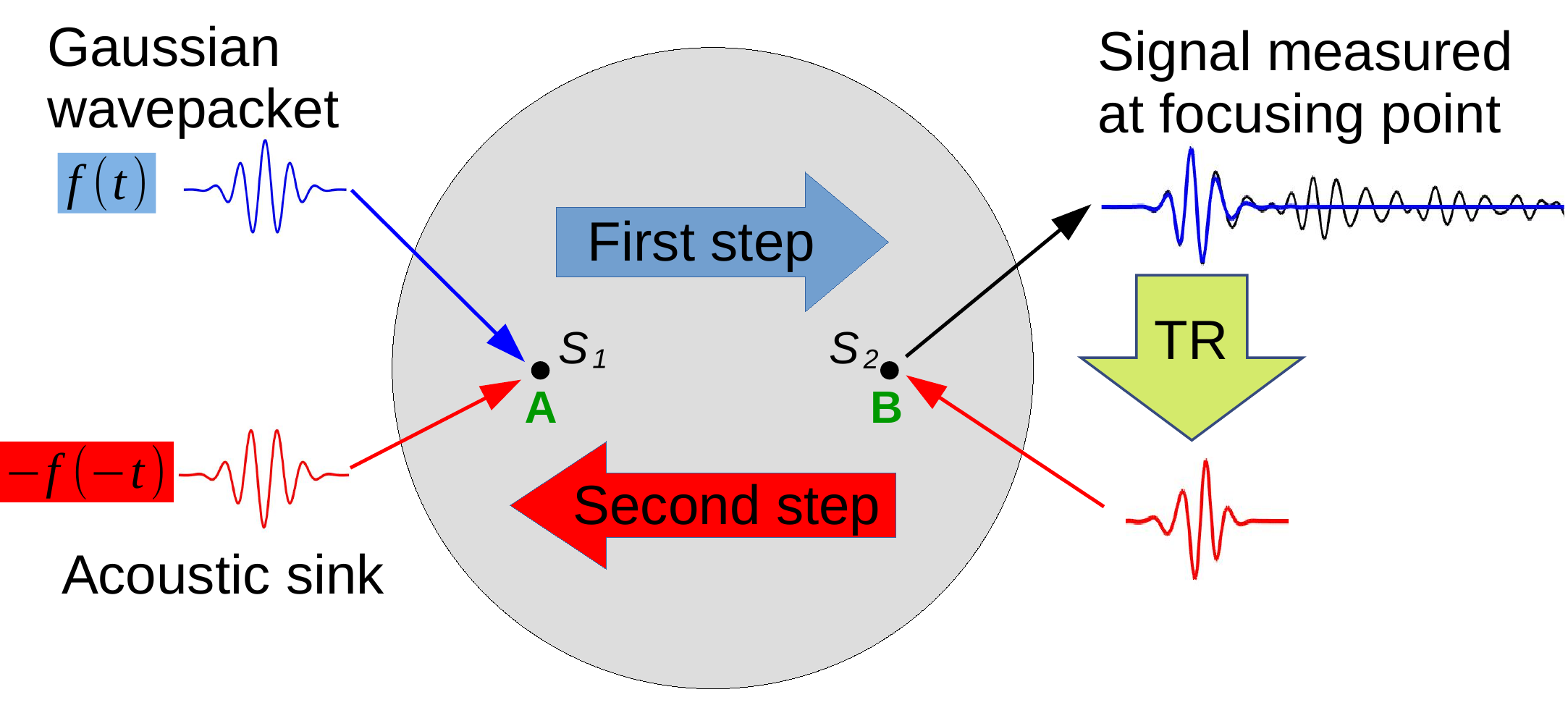}
\end{center}
\caption{  Schematics of the experimental procedure for the TR experiment. In the initial focusing experiment,  $f(t)$ (blue) is emitted at point A and a diffraction limited focusing is observed at point B. The signal recorded at B (black) is shortened to a single pulse (blue) and time-reversed (red). In the second experiment, this signal is emitted from B, while the time-reversed initial source signal is emitted at A to provide an active sink. Sub-lambda focusing is observed at A. } 
\label{fig_TR}
\end{figure}

In order to achieve perfect temporal focusing in addition to spatial refocusing and eliminate the reverberation tail, we propose to use a time-reversed source. To do so, the object and image are now exchanged (see fig.$~$\ref{fig_TR}). In a standard TR experiment with sink \cite{DeRosny2002}, the signal recorded in the initial experiment at point B (fig.$~$\ref{fig_timeFocus}a) would be time-reversed and re-emitted from B by $S_2$, while the TR initial pulse would be emitted by $S_1$, the sink, at A at focusing time to cancel the diverging wave. There, a long sequence of the coda needs to be recorded at B and time-reversed in order to ensure cancel effectively the long reverberation tail. Here, we show that inside the closed Maxwell fisheye, TR of a short time-record corresponding to the first arrival of the pulse (greyed region in Fig.~\ref{fig_timeFocus}a) is enough to fully cancel multiple echoes. By fine tuning the amplitude of $S_1$ and $S_2$, and the time delay between their emission, we are able to perfectly cancel the diverging wave following the focusing, and eliminate multiple echoes, achieving almost perfect temporal pulse compression  as shown in fig.~\ref{fig_timeFocus}c.

To analyse in more details the action of the sink, we perform two independent measurements of the propagation in the lens. By activating the source and the sink separately, we investigate the response to each one. 

\begin{figure}[]
\begin{center}
\includegraphics[width=8.5cm]{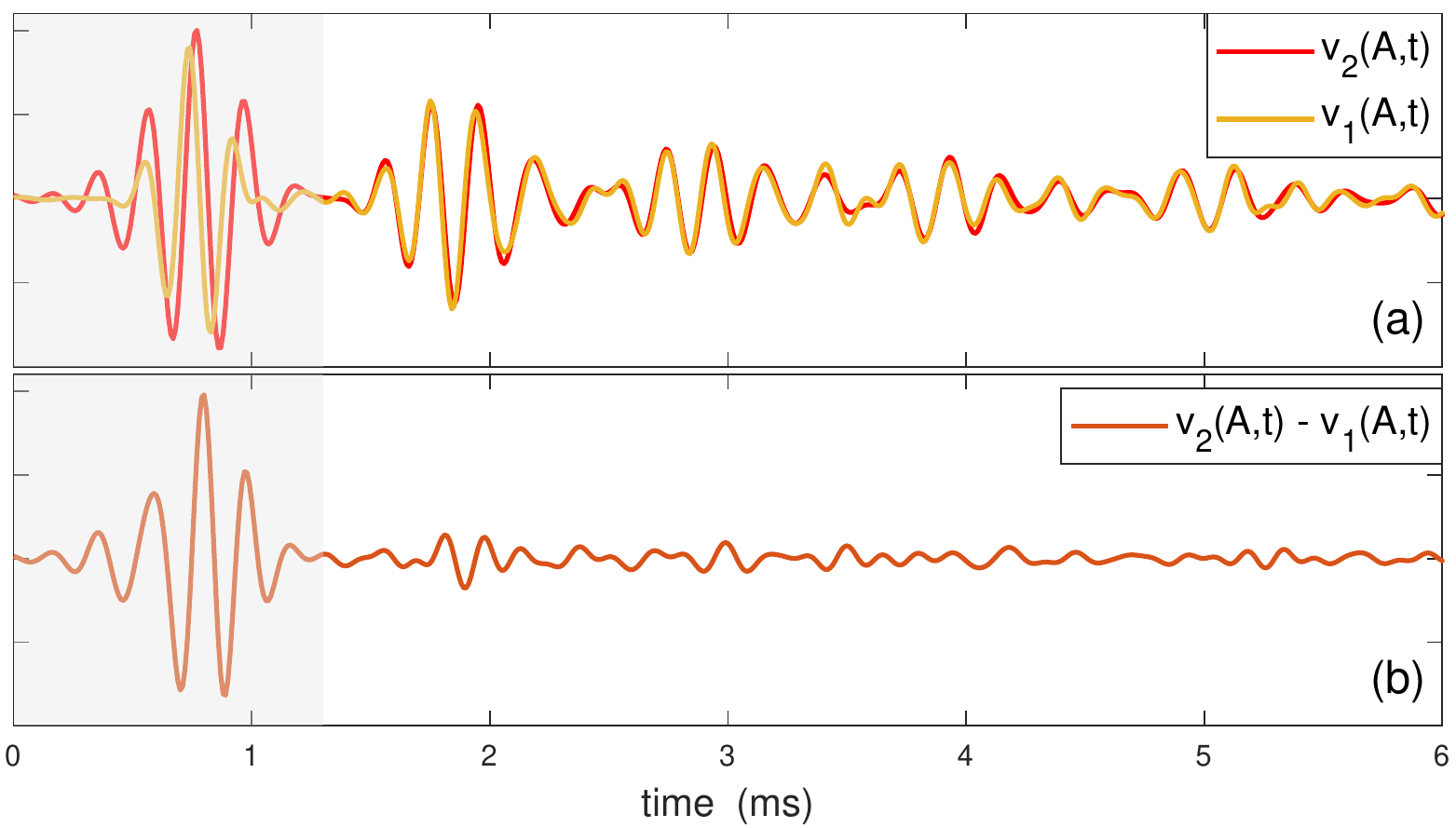}
\end{center}
\mbox{}\vspace{-1cm}
\caption{(a) Velocity signals $v_2$ and $v_1$, generated by source $S_2$ and the sink $S_1$, measured at the focusing point A. The leading pulses (greyed area) and the beginning of very similar long codas are shown. (b) Velocity signal resulting from the two sources acting together (subtracted), measured at the focusing point A, showing that the coda has mostly disappeared. }
\label{fig_antinoise_effect}
\end{figure}

The velocity fields generated respectively by source $S_1$ from point A (initial pulse) and $S_2$ from point B (TR pulse) are measured at the focusing point A, and superimposed in Fig.$~$\ref{fig_antinoise_effect}a. We observe that both signals start with a short pulse (greyed area) followed by multiple echoes. The striking feature is that they share an almost perfectly identical coda after the initial pulse. This explains why the combination of both sources ($S_2$ as emitting source, $S_1$ as sink) allows to cancel with remarkable efficiency the subsequent echoes. This is demonstrated in Fig.$~$\ref{fig_antinoise_effect}b, where the result described in Fig.~\ref{fig_timeFocus}c is reproduced here, simply by superposition of the two signals. Notice that the initial pulses (greyed section) are similar but not identical. They are actually in phase quadrature whereas the codas are perfectly aligned (Fig.$~$\ref{fig_antinoise_effect}a). Therefore, when both sources are emitting together, the tails cancel but the initial pulse remains. The TR source behaves as an optimal drain, emptying all the wavefield from the cavity as a sink. See Supplemental Material at [URL will be inserted by publisher] for a video of showing this result. 

\begin{figure}[]
\begin{center}
\includegraphics[width=8cm]{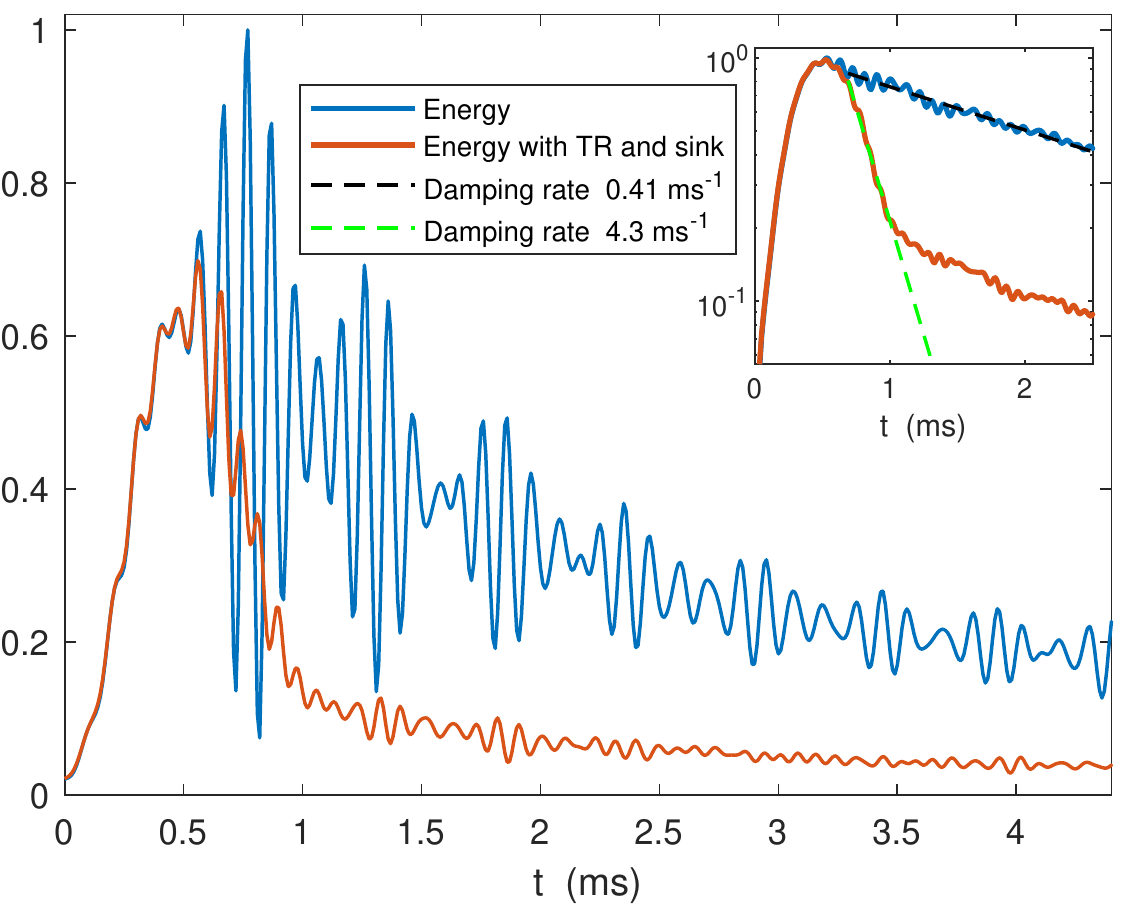}
\end{center}
\mbox{}\vspace{-1cm}
\caption{Time evolution of the instantaneous kinetic energy in the fisheye cavity during the experiment, normalized by the maximal value in the absence of sink. Inset shows the same data averaged over one period in log-scale. Curve fitting allows to evaluate an exponential decay rate for each curve.}
\label{fig_energy}
\end{figure}

The absorption efficiency of the sink is also visible on a global scale, considering the energy stored in the cavity. Figure$~$\ref{fig_energy} displays the time evolution of the total kinetic energy within the plate. It is obtained by integrating over the plate the square of the local velocity field, weighted by the plate thickness. This quantity displays large oscillations as energy conversion between strain and kinetic components occurs. This is especially the case in the absence of sink (blue curve), as an outward wave interferes with the inward wave, creating an important stationary component in the wavefield. Nonetheless, we are here interested in the evolution of the mean value, representative of the total instantaneous energy time evolution. The energy growth seen in fig.$~$\ref{fig_energy} during the first 0.5ms corresponds to the $S_2$ source emission. After that point, the sink demonstrates its astonishing ability to absorb energy from the plate (red curve). A drop of 80\% of the energy level is observed during the following 0.5ms, i.e. for the same duration as the initial pulse. In contrast, the energy drop without the sink is only 12\%. We evaluate the decay rate to be 0.41$~$ms$^{-1}$ without sink, and 4.3$~$ms$^{-1}$ with sink (see inset of Fig.$~$\ref{fig_energy}). Afterward, the energy follows the same trend in both cases. We interpret this full-field cancellation as the result of a complete TR experiment. The TR of both the wavefield and the source allows to reach the same state as before the initial emission, when there was no vibration in the cavity.

Theoretically, a full-360$^{\circ}$ TR mirror or a chaotic cavity would be needed to record all the information encoded in the propagating wavefield, to achieve perfect refocusing of the time-reversed signal at the source position. Working in the Maxwell fisheye lens simplifies the approach, as all the wavefield converges to a single point. Instead of a TR mirror on the boundary of the cavity, a single TR source is enough to time reverse the full wavefield propagating in the cavity. Moreover, it is not necessary to record and emit a long coda signal, in contrast to a standard TR experiment where an infinite time window is theoretically necessary\cite{Draeger1999}. In the transformed medium of the fisheye cavity, this requires a time record as short as the initial pulse.

In the field of active noise control, noise cancellation in a wide room is usually sought through multiple-channel systems \cite{Kuo1999}. Thanks to the fisheye cavity imaging property, here we obtain a quantitative noise cancellation with only one source. 
In their analysis of the fisheye resolution, Tyc and Danner showed theoretically how to build an optimal absorbing source, taking advantage of the spherical modal decomposition \cite{Tyc2014}. We propose here a practical realisation of this possibility, but without using an a priori calculation of the desired signal. We show that the experimental time-reversal procedure itself indicates how to design a sink to achieve perfect energy draining.

This approach can be an alternative to other methods, such as elastic "black holes" \cite{krylov2004new,kralovic2007damping,bowyer2014damping,tang2017enhanced,deng2019ring, Pelat2020}, or absorbing systems using electrical shunts \cite{ELLIOTT2014}. The interest of a lens to concentrate energy and improve local harvesting has already been investigated \cite{Yi2017}. The efficiency displayed by a fisheye lens combined with time-reversal and sink suggests this is a method with high potential for energy harvesting of plate vibrations. 
\textcolor{black}{Thus, our work paves the way to absolute instruments for waves in both space and time domains making use of transformed media and time reversal.}

P.S. is thankful to the Agence Nationale de la Recherche support under grant ANR PLATON (No. 12-BS09-003-01), the LABEX WIFI (Laboratory of Excellence within the French Program Investments for the Future) under reference ANR-10-IDEX-0001-02 PSL* and the Groupement de Recherche 3219 MesoImage. 
Figure$~$\ref{fig_intro}$~$(a-b) was created by company 3DBM. 

\bibliography{Fisheye_Biblio}

\end{document}

\title{Subwavelength pulse focusing and perfect absorption in the Maxwell fisheye \\ Supplementary material}
\maketitle

In this work, we assumed that the flexural vibrations on an elastic isotropic homogeneous spherical shell of small thickness can be mapped on those of a thin elastic isotropic homogeneous plate with a varying thickness. This correspondence can be deduced as follows. We first note that the governing Navier equation for elastodynamic waves propagating within the spherical shell can be written as follows in
terms of the displacement field ${\bf u}=(u_1,u_2,u_3)(x_1,x_2,x_3)$:
\begin{equation}
\frac{E(1-\nu)}{(1+\nu)(1-2\nu)}\nabla(\nabla\cdot{\bf u})
-\frac{E}{2(1+\nu)}\nabla\times(\nabla\times{\bf u})
= \rho\frac{\partial^2}{\partial t^2}{\bf u}
\label{eqsupp1}
\end{equation}
where $E$ where $E$ is the Young’s modulus of the shell, $\nu$ is its
Poisson’s ratio, and $\rho$ is its density. Moreover, $\nabla=(\partial/\partial x_1,\partial/\partial x_2,\partial/\partial x_3)^T$. 

We can then make use of the von Karmann theory of thin elastic plates since the thickness of the spherical shell is much smaller than the wavelength elastic vibrations. For this, we note that the Navier equation (\ref{eqsupp1}) can be simplified using the Helmholtz decomposition theorem with a scalar acoustic
potential $\Phi$ and a vector elastic potential $\Psi=(\Psi_1,\Psi_2,\Psi_3)$.
These can be expressed in terms of the theta function $\Theta$ 
 in the
following manner:
\begin{equation}
\begin{array}{cc}
\Phi(x_1,x_2,x_3,t)&=\phi(t)\Delta\log(\Theta(x_1,x_2,x_3)) \; , \\
\Psi_j(x_1,x_2,x_3,t)&=\psi_j(t)\Delta\log(\Theta(x_1,x_2,x_3)) \; , \\
\label{eqsupp2}
\end{array}
\end{equation}
where $j = 1,2,3$.

Let us now assume some time-harmonic vibrations. It is possible to show that the theta function
is a solution of the von Karmann equation \cite{graff2012wave}
\begin{equation}
\nabla\cdot\{\xi^{-1}\nabla[\Delta\nabla\cdot(\xi^{-1}\nabla\Theta)]\}-\Lambda^{-1}\beta_0^4\Theta=0 \; ,
\label{eqsupp3}
\end{equation}
where $\nabla=(\partial/\partial x_1,\partial/\partial x_2,\partial/\partial x_3)^T$, $\Delta=\nabla\cdot\nabla$ and $\Lambda=\rho^{-1}$, $\xi=E^{-1/2}$, with $E$ the Young modulus of the composite constituting the shell and $\beta_0^4=\omega^2\rho_0h_0/D_0$, with $D_0$ the rigidity of the shell, $\rho_0$ its density, $h_0$ its thickness, and $\omega$ the wave frequency.

This sixth order partial differential equation has good properties, since it retains its form under geometric transform \cite{munteanu2011three,munteanu2014new}, and the transformed equation reads as
\begin{equation}
\nabla'\cdot\{\underline{\underline{\xi}}_{t}^{-1}\nabla'[\Delta'_{33}\nabla'\cdot(\underline{\underline{\xi}}_{t}^{-1}\nabla'\Theta)]\}-\Lambda_{33}^{-1}\beta_0^4\Theta=0 \; ,
\label{eqsupp4}
\end{equation}
with the transformed parameters $\underline{\underline{\xi}}_{t}^{-1}$ and $\Lambda_{33}^{-1}$ being defined as
the upper Block diagonal part of the rank 2 tensor $\underline{\underline{\xi}}^{-1}$ and the third diagonal coefficient of the rank 2 tensor $\underline{\underline{\Lambda}}^{-1}$ given by 
\begin{equation}
\underline{\underline{\xi}}^{-1}=J^T\xi^{-1}J \det(J^{-1}) \; , \;
\underline{\underline{\Lambda}}^{-1}=J^T\Lambda^{-1}J \det(J^{-1})
\; ,
\label{eqsupp5}
\end{equation}
where $J$ is the Jacobian matrix of the coordinate change $(x_1,x_2,x_3)\longrightarrow (X_1,X_2,X_3)$. Moreover, $\nabla'=(\partial/\partial X_1,\partial/\partial X_2,\partial/\partial X_3)^T$, $\Delta'_{33}=\partial^2/\partial X_3^2$.

We now wish to apply a transform that gets rid of the third coordinate: $(x_1,x_2,x_3)\longrightarrow (X_1,X_2)$. This is nothing but a projection. We note that the stereographic projection of the sphere onto the plane is given by
\begin{equation}
X_1=\frac{x_1}{1-x_3/R} \; , \; X_2=\frac{x_2}{1-x_3/R} \; ,
\label{eqsupp6}
\end{equation}
and making use of the inverse formula
\begin{equation}
x_1=\frac{2X_1}{1+(r/R)^2} \; , \; x_2=\frac{2X_2}{1+(r/R)^2} \; , \; x_3=R\frac{(r/R)^2-1}{1+(r/R)^2} \; ,
\label{eqsupp7}
\end{equation}
where $r=\sqrt{X_1^2+X_2^2}$ and $R$ is the radius of the thin spherical shell, we deduce that the transformed equation for the thin plate in coordinates $(X_1,X_2)$ reduces to the fourth order Kirchhoff-Love partial differential equation
\begin{equation}
\Delta'\Delta' W-h(r)\beta_0^4W=0 \; ,
\label{eqsupp8}
\end{equation}
where $\Delta'=\partial^2/\partial X_1^2+\partial^2/\partial X_2^2$, with $W(X_1,X_2)$ the out-of-plane displacement along $X_3$. Besides,
$h(r)=(1+(r/R)^2)^2$. This transfomed equation corresponds to that used in \cite{Lefebvre2015}.
\bibliography{Fisheye_Biblio}